\documentstyle[11pt,epsfig]{article}
\begin{document}

\title{The Decays of $B^{+}\rightarrow{\bar{D}^0}+D^{+}_{sJ}(2S)$
and $B^{+}\rightarrow\bar{D}^{0}+D^{+}_{sJ}(1D)$} \vspace{10mm}

\author{Guo-Li Wang\footnote{gl\_wang@hit.edu.cn},~~Jin-Mei Zhang~~and
Zhi-Hui Wang \\
\\
{\it \small  Department of Physics, Harbin Institute of
Technology, Harbin 150001, China} }
\date{}
\maketitle

\baselineskip=24pt

\begin{quotation}

\vspace*{1.5cm}
\begin{center}
  \begin{bf}
  ABSTRACT
  \end{bf}
\end{center}

\vspace*{0.5cm} \noindent We analyzed the decays of
$B^{+}\rightarrow {\bar {D}^0}+D^{+}_{sJ}(2S)$ and
$B^{+}\rightarrow\bar D^{0}+D^{+}_{sJ}(1D)$ by naive factorization
method and model dependent calculation based on the Bethe-Salpeter
method, the branching ratios are $Br({\bar
{D}^0}D^{+}_{sJ}(2S))=(0.72\pm0.12)\%$ and $Br({\bar
{D}^0}D^{+}_{sJ}(1D))=(0.027\pm0.007)\%$. The branching ratio of
decay $B^{+}\rightarrow {\bar {D}^0}D^{+}_{sJ}(2S)\rightarrow {\bar
{D}^0}D^0 K^+$ consist with the data of Belle Collaboration, so we
conclude that the new state $D^{+}_{sJ}(2700)$ is the first excited
state $D^{+}_{sJ}(2S)$.

\end{quotation}

\newpage
\setcounter{page}{1}

BABAR Collaboration reported the observation of a new charmed meson
called $D^{+}_{sJ}(2690)$ with a mass $2688\pm 4\pm 3$ MeV and with
a broad width $\Gamma = 112\pm 7\pm 36$ MeV \cite{BABAR}, Belle
Collaboration also reported a charmed state $D^{+}_{sJ}(2700)$,
$M=2715\pm 11^{+11}_{-14}$ MeV with width $\Gamma=115\pm
20^{+36}_{-32}$ MeV \cite{Belle1}, later modified to $M=2708\pm
9^{+11}_{-10}$ MeV and $\Gamma=108\pm 23^{+36}_{-31}$ MeV
\cite{Belle2}. Most authors believe that these two states should be
one state, and the most possible candidate is the $2S$ or the $S-D$
($2~^3S_1-1~^3D_1$) mixing $c\bar s$ $1^-$ state
\cite{close,zhu,rosner,fazio,matsuki,dmli,zhao,li}.

Recently, we have analyzed the decays of the $D^{+}_{sJ}(2S)$ and
 $D^{+}_{sJ}(1D)$, we concluded that one can not distinguish them
from their decays, because they have the similar decay channels and
the corresponding decay widths are comparable \cite{2700}. In this
paper, we give the calculations of the decays $B^{+}\rightarrow
{\bar {D}^0}+D^{+}_{sJ}(2S)$ and $B^{+}\rightarrow\bar
D^{0}+D^{+}_{sJ}(1D)$, we find that we can separate them, and
according to the current experimental data of Belle Collaboration,
we conclude that the new state $D^{+}_{sJ}(2700)$ is the first
excited state $D^{+}_{sJ}(2S)$. These channels have also been
considered in literatures, for example, Close et al
\cite{close,close2} give branching ratios and possible mixing of
$2S$ and $1D$; Colangelo et al \cite{fazio} assume the new state is
the $2S$ state, and extract the decay constant.

The decay amplitude for $B^{+}\rightarrow {\bar {D}^0}+D^{+}_{sJ}$
can be described in the naive factorization approach:
\begin{eqnarray}
T=\frac{G_F}{\sqrt{2}}V_{cs}V_{cb}a_1\langle
D_{_{sJ}}^{+}|J_{\mu}|0\rangle\langle{\bar
D^0}|J^{\mu}|B^+\rangle,\label{eq1}
\end{eqnarray}
where the CKM matrix element $V_{cs}=0.97334$ and $V_{cb}=0.0415$
\cite{pdg}, since we focus on the difference between
$D^{+}_{sJ}(2S)$ and $D^{+}_{sJ}(1D)$, not on the careful study, we
have chosen $a_1=c_1+\frac{1}{3}c_2=1$, where $c_1$ and $c_2$ are
the Wilson coefficients \cite{wilson}. We also delete other higher
order contributions like the contributions from penguin operators.
And
\begin{eqnarray}\langle
D_{sJ}^{+}|J_{\mu}|0\rangle=iF^{*}_V
M_{{D_{_{sJ}}^+}}\epsilon^{\lambda}_{\mu}~,\label{eq2}\end{eqnarray}
$F_V$ and $\epsilon^{\lambda}_{\mu}$ are the decay constant and
polarization vector of the meson $D_{_{sJ}}^+$, respectively.

We have solved the exact instantaneous Bethe-Salpeter equations
\cite{BS} (or the Salpeter equations \cite{salp}) for $1^{-}$
states, the general form for the relativistic Salpeter wave function
of vector $1^{-}$ state can be written as \cite{changwang,wang1}:
$$\varphi_{1^{-}}^{\lambda}(q_{\perp})=
q_{\perp}\cdot{\epsilon}^{\lambda}_{\perp}
\left[f_1(q_{\perp})+\frac{\not\!P}{M}f_2(q_{\perp})+
\frac{{\not\!q}_{\perp}}{M}f_3(q_{\perp})+\frac{{\not\!P}
{\not\!q}_{\perp}}{M^2} f_4(q_{\perp})\right]+
M{\not\!\epsilon}^{\lambda}_{\perp}f_5(q_{\perp})$$
\begin{equation}+
{\not\!\epsilon}^{\lambda}_{\perp}{\not\!P}f_6(q_{\perp})+
({\not\!q}_{\perp}{\not\!\epsilon}^{\lambda}_{\perp}-
q_{\perp}\cdot{\epsilon}^{\lambda}_{\perp})
f_7(q_{\perp})+\frac{1}{M}({\not\!P}{\not\!\epsilon}^{\lambda}_{\perp}
{\not\!q}_{\perp}-{\not\!P}q_{\perp}\cdot{\epsilon}^{\lambda}_{\perp})
f_8(q_{\perp}),\label{eq13}
\end{equation}
where the $P$, $q$ and ${\epsilon}^{\lambda}_{\perp}$ are the
momentum, relative inner momentum and polarization vector of the
vector meson, respectively; $f_i(q_{\perp})$ is the function of
$-q_{\perp}^2$, and we have used the notation $q^{\mu}_{\perp}\equiv
q^{\mu}-(P\cdot q/M^{2})P^{\mu}$ (which is $(0,~\vec q)$ in the
center of mass system).

The $8$ wave functions $f_i$ are not independent, there are only $4$
independent wave functions \cite{wang1}, and we have the relations
$$f_1(q_{\perp})=\frac{\left[q_{\perp}^2 f_3(q_{\perp})+M^2f_5(q_{\perp})
\right] (m_1m_2-\omega_1\omega_2+q_{\perp}^2)}
{M(m_1+m_2)q_{\perp}^2},~~~f_7(q_{\perp})=\frac{f_5(q_{\perp})M(-m_1m_2+\omega_1\omega_2+q_{\perp}^2)}
{(m_1-m_2)q_{\perp}^2},$$
$$f_2(q_{\perp})=\frac{\left[-q_{\perp}^2 f_4(q_{\perp})+M^2f_6(q_{\perp})\right]
(m_1\omega_2-m_2\omega_1)}
{M(\omega_1+\omega_2)q_{\perp}^2},~~~f_8(q_{\perp})=\frac{f_6(q_{\perp})M(m_1\omega_2-m_2\omega_1)}
{(\omega_1-\omega_2)q_{\perp}^2}.$$

In our method, the $S-D$ mixing automatically exist in the wave
function of $1^-$ state, because we give the whole wave function Eq.
(\ref{eq13}) which is $J^P=1^-$, but some of the wave function are
$S$ wave, some are $D$ wave, for example, see Figure 1-3, we show
wave functions. One can see that, for $1S$ and $2S$, the wave
function $f_5$ and $f_6$ are dominate, and they are $S$ wave, but
there is little $D$ wave mixing in these two states which come from
the terms $f_3$ and $f_4$. But for the third state, we labeled as
$1D$ state, the terms $f_3$ and $f_4$ are dominate, but these two
terms are not pure $D$ wave, they will give contribute as a $S$ wave
\cite{changwang}. So we conclude that the $1S$ and $2S$ states are
$S$ wave dominate states, mixed with a little $D$ wave (come from
the terms $f_3$ and $f_4$), while the $1D$ state is a $D$ wave
dominate state ($f_3$ and $f_4$), mixed with a valuable part of $S$
wave (still come from the terms $f_3$ and $f_4$).

\begin{figure}
\centering
\includegraphics[width=0.65\textwidth]{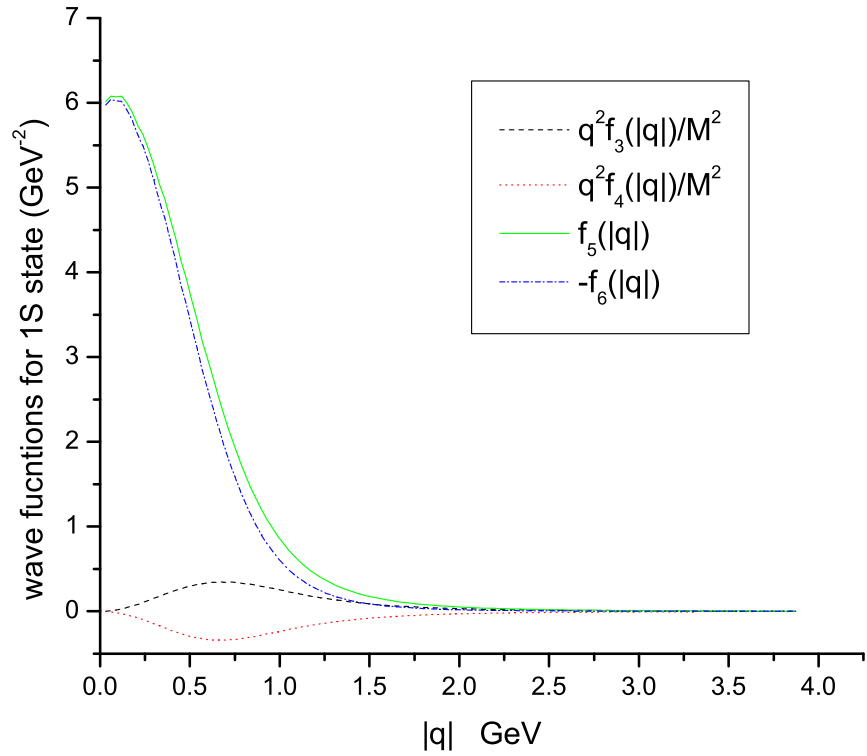}
\caption{\label{fig:1s}wave functions of $D_{s}^{*}(1S)$.}
\end{figure}

\begin{figure}
\centering
\includegraphics[width=0.65\textwidth]{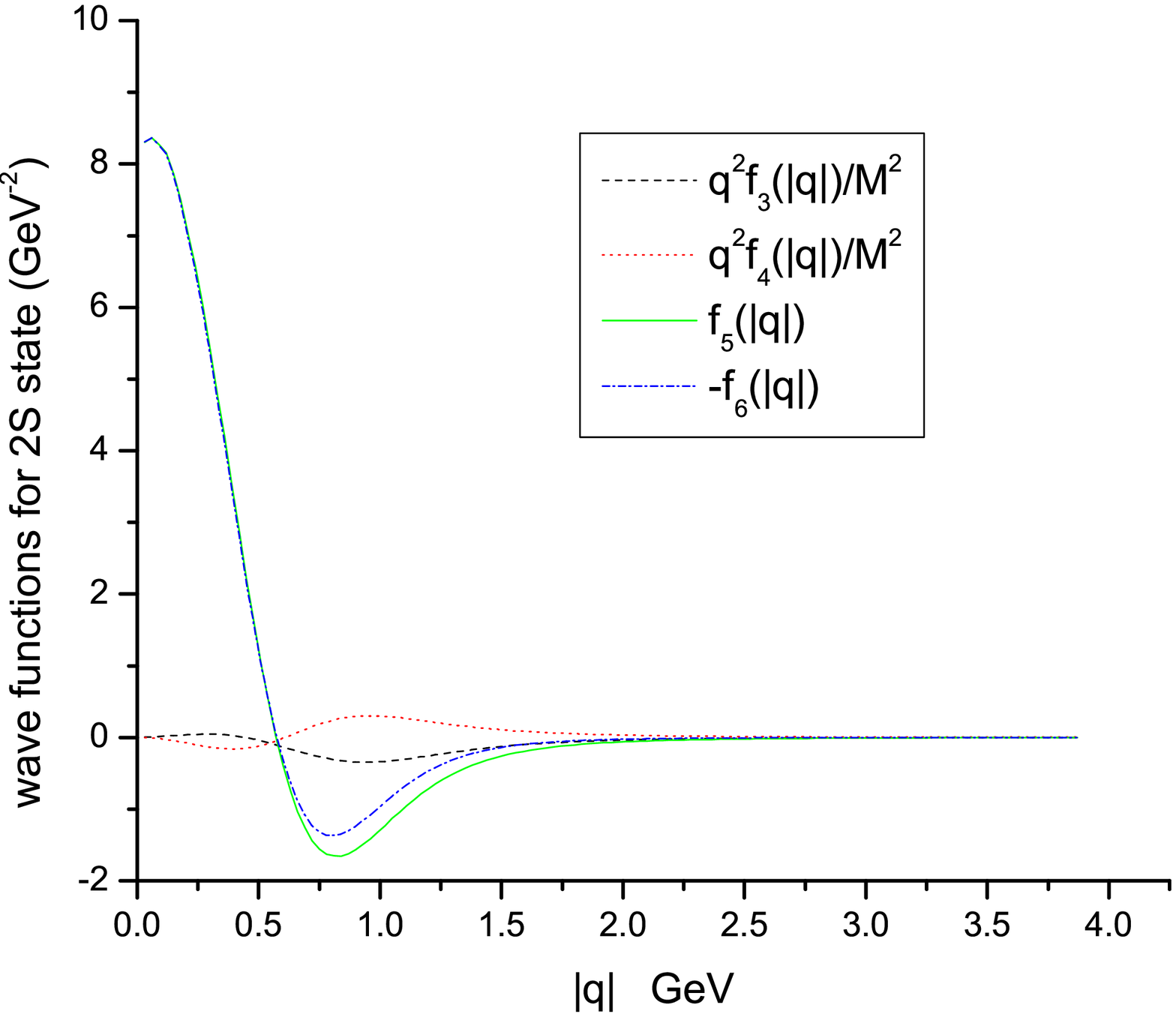}
\caption{\label{fig:2s}wave functions of $D_{s}^{*}(2S)$.}
\end{figure}

\begin{figure}
\centering
\includegraphics[width=0.65\textwidth]{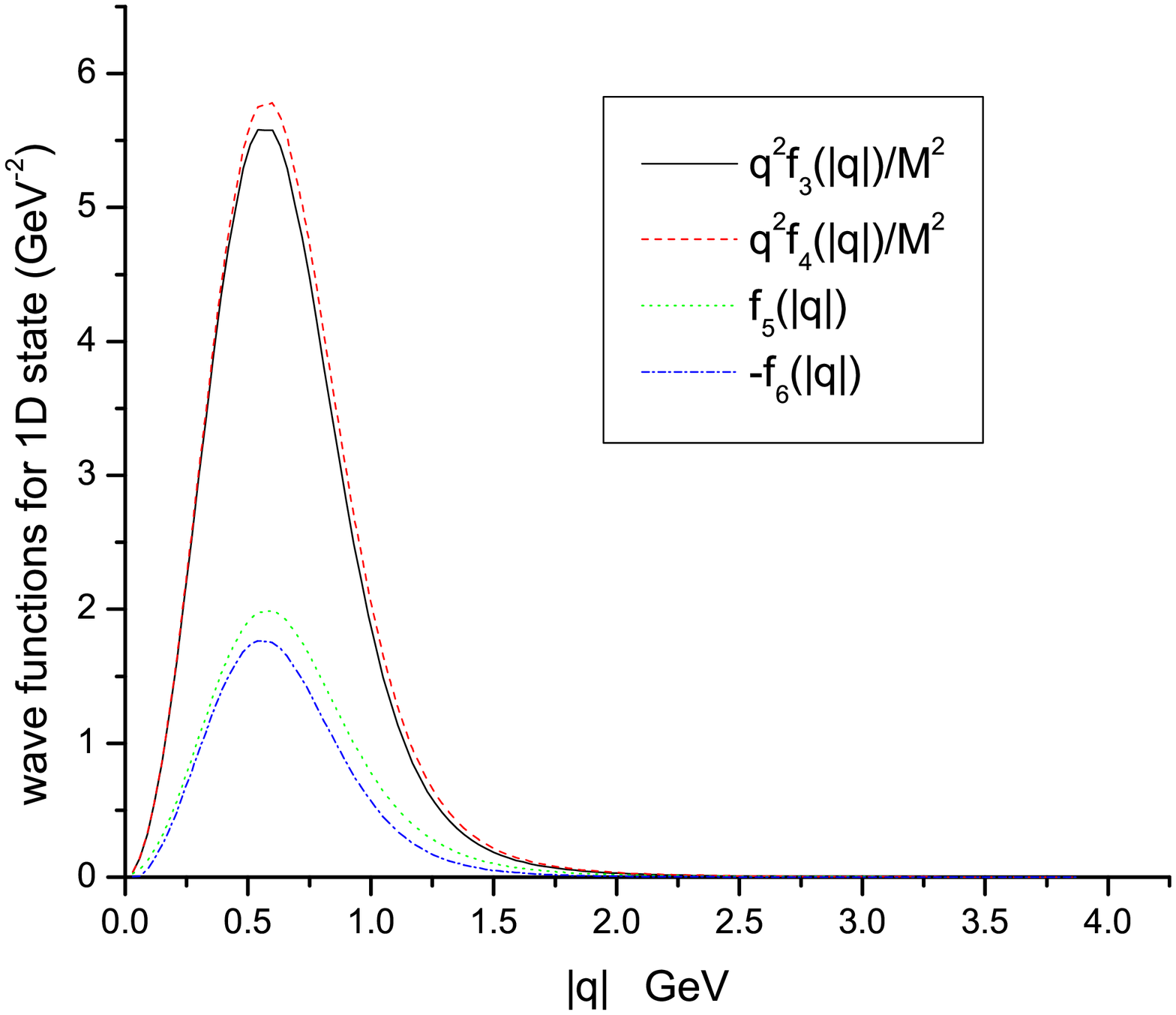}
\caption{\label{fig:1d}wave functions of $D_{s}^{*}(1D)$.}
\end{figure}

For $c\bar s$ vector $1^-$ state, our mass prediction for the first
radial excited $2 S$ state is $2673$ MeV, and for $1 D$, our result
is $2718$ MeV \cite{wang1}, so there are two states around $2700$
MeV.

In Ref. \cite{wang1}, we only give the leading order calculation for
decay constant, which is $ F_{V} = 4\sqrt{N_c} \int \frac{d{\vec
q}}{(2\pi)^3} f_{5}({\vec q})$, the whole equation should be
\begin{equation} F_{V} = 4\sqrt{N_c} \int \frac{d{\vec q}}{(2\pi)^3}
(f_{5}-\frac{{\vec q}^2f_{3}}{3M^2}),\end{equation} and our results
of the decay constants for vector $c\bar s$ system are:
$$F_V(1S)=353\pm 21~~\rm MeV,$$
$$F_V(2S)=295\pm 13~~\rm MeV,$$
$$F_V(1D)=57.1\pm 5.1~~\rm MeV.$$
Where the uncertainties are given by varying all the input
parameters simultaneously within $\pm5\%$ of the central values in
our model, and we will calculate all the uncertainties this way in
this letter. The center values of $S$ waves are little smaller than
the estimates in Ref. \cite{wang1}, where $F_V(1S)=375\pm 24~\rm
MeV$ and $F_V(2S)=312\pm 17~\rm MeV$, but the value of $D$ wave
decay constant is much smaller than the predict in Ref. \cite{wang1}
where we did not shown it, this is because we should not ignore the
contribution from the term of $f_3$ when consider a $D$ wave state.
Our estimate of $F_V(1S)=353\pm 23~\rm MeV$ is close to the newer
result $F_{D^{*}_{s}}=268\sim290~\rm MeV$ by Choi \cite{choi}. Our
estimate of $F_V(2S)=312\pm 17~\rm MeV$ is a little larger than the
estimate $F_V(2S)=243\pm 41~\rm MeV$ in Ref. \cite{fazio}, where
they extracted it from the decay $B^{+}\rightarrow {\bar
{D}^0}+D^{+}_{sJ}(2700)$ assuming the new state $D^{+}_{sJ}(2700)$
is the first radial excited state $D^{+}_{sJ}(2S)$.

According to the Mandelstam formalism \cite{mandelstam}, at the
leading order, the transition amplitude for $B^+ \rightarrow
\bar{D^0}$ can be written as \cite{2632}:
\begin{eqnarray}
\langle{\bar D^0}|J^{\mu}|B^+\rangle= \int\frac{d{\vec
q}}{(2\pi)^3}Tr\left[ \bar{\varphi}^{++}_{_{\bar D^{0}}}({ \vec
q}-\frac{m_u}{m_c +m_u}{\vec r })\frac{\not\!P}{M}
{\varphi}^{++}_{_{B^{+}}}({\vec q})\gamma^{\mu}(1-\gamma_{5})
\right],\label{eq3}
\end{eqnarray}
where $\vec r$ is the recoil three dimensional momentum of the final
state $\bar D^0$ meson, ${\varphi}^{++}$ is called the positive
energy wave function, and $\bar{\varphi}^{++}_{_{\bar
D^{0}}}=\gamma_{0}({\varphi}^{++}_{_{\bar D^{0}}})^{+}\gamma_{0}$.

The wave function forms for pseudoscalar $B^+$ and $\bar{D^0}$ are
similar, for example, the wave function for $B^+$ can be written as
\cite{cskimwang}
$$
\varphi^{++}_{_{B^{+}}}(\stackrel{\rightarrow}{q})=
\frac{M_{{B^+}}}{2}\left(\varphi_1(\stackrel{\rightarrow}{q})
+\varphi_2(\stackrel{\rightarrow}{q})\frac{m_{u}+m_{b}}{\omega_{u}+\omega_{b}}\right)
$$\begin{equation}\times\left[
\frac{\omega_{u}+\omega_{b}}{m_{u}+m_{b}}+{\gamma_{_0}}-\frac{{\not\!
\vec q}(m_{u}-m_{b})}
{m_{b}\omega_{u}+m_{u}\omega_{b}}+\frac{{\not\! \vec
q}\gamma_{_0}(\omega_{u}
+\omega_{b})}{(m_{b}\omega_{u}+m_{u}\omega_{b})}\right]\gamma_{_5}\;,
\end{equation}
where $\omega_{u}=\sqrt{m_{u}^{2}+{\vec q}^{2}}$ and
$\omega_{b}=\sqrt{m_{b}^{2}+{\vec q}^{2}}$; $\varphi_1(\vec{q})$,
$\varphi_2(\vec{q})$ are the radial part wave functions, and their
numerical values can be obtained by solving the full Salpter
equation of $0^{-}$ state \cite{cskimwang}.

The decay width is:
$$\Gamma=\frac{1}{8\pi}\frac{|\vec{p}_{_{f_2}}|}{M^2_B}|T|^2$$
\begin{eqnarray}
=\frac{1}{8\pi}\frac{|\vec{p}_{_{f_2}}|}{M^2_B}
\frac{G^2_{F}V^2_{cs}V^2_{bc}}{2}F^2_VM^2_{f_2}X,\label{eq37}
\end{eqnarray}
where $\vec{p}_{_{f_2}}$ and $M_{f_2}$ are the three dimensional
momentum and mass of the final new state $D^{*+}_{sJ}$. $M^2_{f_2}$
come from the definition of decay constant in Eq.(\ref{eq2}), but
the square of polarization vector and transition amplitude of
Eq.(\ref{eq3}) are symbolized as $X$.

So our result are:
\begin{eqnarray}
\Gamma(B^{+}\rightarrow {\bar
{D}^0}D^{*+}_{s}(2S))=F^2_V(2S)\times(3.50\pm0.38)\times
10^{-14}~\rm GeV\label{eq17},
\end{eqnarray}
\begin{eqnarray}
\Gamma(B^{+}\rightarrow {\bar
{D}^0}D^{*+}_{s}(1D))=F^2_V(1D)\times(3.25\pm0.30)\times
10^{-14}~\rm GeV. \label{eq27}
\end{eqnarray}
If we ignore the mass difference between $D^{*+}_{s}(2S)$ and
$D^{*+}_{s}(1D)$, and use $2700$ MeV as input, the results become
\begin{eqnarray}
\Gamma(B^{+}\rightarrow {\bar
{D}^0}D^{*+}_{s}(2S))=F^2_V(2S)\times(3.36\pm0.25)\times
10^{-14}~\rm GeV\label{eq117},
\end{eqnarray}
\begin{eqnarray}
\Gamma(B^{+}\rightarrow {\bar
{D}^0}D^{*+}_{s}(1D))=F^2_V(1D)\times(3.36\pm0.25)\times
10^{-14}~\rm GeV. \label{eq217}
\end{eqnarray}
So the difference between this two channel mainly come from the
difference of decay constants. Then our predictions of branching
ratios are:
\begin{eqnarray}
Br(B^{+}\rightarrow {\bar {D}^0}D^{*+}_{s}(2S))=(0.72\pm0.12)\%,
\label{eq7}
\end{eqnarray}
\begin{eqnarray}
Br(B^{+}\rightarrow {\bar {D}^0}D^{*+}_{s}(1D))=(0.027\pm0.007)\%.
\label{eq8}
\end{eqnarray}

In Ref. \cite{2700}, we have calculated the main decay channels of
$D^{+}_{sJ}(2S)$ and $D^{+}_{sJ}(1D)$, and we have the following
estimates:
$$Br(D^{+}_{sJ}(2S)\rightarrow D^{0}K^{+})= 0.20\pm 0.03$$
and
$$Br(D^{+}_{sJ}(1D)\rightarrow D^{0}K^{+})= 0.32\pm0.04,$$
so we obtain: \begin{equation}Br(B^{+}\rightarrow {\bar
{D}^0}D^{+}_{sJ}(2S))\times Br(D^{+}_{sJ}(2S)\rightarrow
D^{0}K^{+})= (1.4\pm 0.5)\times 10^{-3}\end{equation} and
\begin{equation}Br(B^{+}\rightarrow {\bar
{D}^0}D^{+}_{sJ}(1D))\times Br(D^{+}_{sJ}(1D)\rightarrow
D^{0}K^{+})= (0.9\pm0.3)\times 10^{-4}.\end{equation} Our estimate
of $Br(D^{+}_{sJ}(2S)\rightarrow D^{0}K^{+})\simeq 0.20$ is larger
than than the estimate $0.11$ in Ref. \cite{close} and the estimate
$0.05$ in Ref. \cite{zhu}, if we use their value as input, we will
obtain a smaller value of $Br(B^{+}\rightarrow {\bar
{D}^0}D^{+}_{sJ}(2S))\times Br(D^{+}_{sJ}(2S)\rightarrow
D^{0}K^{+})$. But our estimate of $Br(D^{+}_{sJ}(1D)\rightarrow
D^{0}K^{+})\simeq 0.32$ consist with the value $0.34$ in Ref.
\cite{zhu}. One can also see that, the final branching ratios depend
strongly on the decay constants, but at this time few papers have
calculated the values of decay constants for $D^{+}_{sJ}(2S)$ and
$D^{+}_{sJ}(1D)$. In Ref. \cite{fazio}, they assumed that the new
state $D^{+}_{sJ}(2700)$ is $D^{+}_{sJ}(2S)$, and they extracted
decay constant of $D^{+}_{sJ}(2S)$: $F_{D^{+}_{sJ}(2S)}=243\pm41$
MeV.

The Belle Collaboration have the data \cite{Belle2,pdg}
$$Br(B^{+}\rightarrow {\bar
{D}^0}D^{+}_{sJ}(2700))\times Br(D^{+}_{sJ}(2700)\rightarrow
D^{0}K^{+})=(1.13^{~+0.26}_{~-0.36})\times 10^{-3}.$$ Because we
only calculated the decay widths of six main channels for
$D^{+}_{sJ}(2S)$ ($D^{+}_{sJ}(1D)$), and based on the summed width
of these six channels, not full width, we give a relative larger
branching ratio $Br(D^{+}_{sJ}(2S)\rightarrow D^{0}K^{+})$ and
$Br(D^{+}_{sJ}(1D)\rightarrow D^{0}K^{+})$, the real branching
ratios should be smaller than our estimates, so our estimate of
$B^+$ decay to $D^{+}_{sJ}(2S)$ is close to the data, while the
estimate of $B^+$ decay to $D^{+}_{sJ}(1D)$ is much smaller than the
data, then we can draw a conclusion that the new state
$D^{+}_{sJ}(2700)$ is $D^{+}_{sJ}(2S)$.

We have another method to estimate the branching ratio, because the
mass of $B^+$ is much heavier than the mass of ${\bar {D}^0}$, and
the mass of $D^{+}_{sJ}(2700)$ is close to the mass of $D^{*+}_{s}$,
as a rough estimate, we ignore the mass difference of
$D^{+}_{sJ}(2700)$ and $D^{*+}_{s}(2112)$, from Eq.(\ref{eq1}) and
Eq.(\ref{eq2}), then we have
\begin{equation}
\frac{Br(B^{+}\rightarrow {\bar
{D}^0}D^{+}_{sJ}(2S))}{Br(B^{+}\rightarrow {\bar
{D}^0}D^{*+}_{s}(1S))}\simeq\frac{F^2_V(2S) }{F^2_V(1S) },
\end{equation}
and from Particle Data Group \cite{pdg}, the branching ratio of
$B^{+}\rightarrow {\bar {D}^0}+D^{+}_{sJ}(1S)$:
$$Br(B^{+}\rightarrow {\bar
{D}^0}D^{*+}_{s}(1S))=(7.8\pm 1.6)\times 10^{-3}.$$ So we have:
\begin{equation}Br(B^{+}\rightarrow {\bar
{D}^0}D^{*+}_{s}(2S))\simeq\frac{F^2_V(2S)}{F^2_V(1S)}\times(7.8\pm
1.6)\times 10^{-3}\simeq (5.4\pm 1.7)\times 10^{-3},\end{equation}
\begin{equation}Br(B^{+}\rightarrow {\bar
{D}^0}D^{*+}_{s}(1D))\simeq\frac{F^2_V(1D)}{F^2_V(1S)}\times(7.8\pm
1.6)\times 10^{-3} \simeq(2.0\pm 1.0)\times 10^{-4}.\end{equation}
This rough estimate results are very close to our calculations
(Eq.(\ref{eq7}) and Eq.(\ref{eq8})), so we have the same conclusion
that $D^{+}_{sJ}(2700)$ is $D^{+}_{sJ}(2S)$.

In Ref. \cite{2700}, we estimate the full widths of $D^{+}_{sJ}(2S)$
and $D^{+}_{sJ}(1D)$ by six main decay channels, the estimated full
widths are $46.4 \pm 6.2$ MeV for $D^{+}_{sJ}(2S)$, $73.0 \pm 10.4$
MeV for $D^{+}_{sJ}(1D)$, comparing with experimental data, $\Gamma
= 112\pm 7\pm 36$ MeV for $D^{+}_{sJ}(2690)$ and $\Gamma=108\pm
23^{+36}_{-31}$ MeV for $D^{+}_{sJ}(2700)$, there is still the
possible that there are two states around $2700$ MeV, one is the $S$
wave dominate $D^{+}_{sJ}(2S)$, the other is $D$ wave dominate
$D^{+}_{sJ}(1D)$.

As summary, from the decays $B^{+}\rightarrow {\bar
{D}^0}+D^{+}_{sJ}(2S)$ and $B^{+}\rightarrow\bar
D^{0}+D^{+}_{sJ}(1D)$, we conclude that the new state
$D^{+}_{sJ}(2700)$ is the first radial excited state
$D^{+}_{sJ}(2S)$, and there may exist another state around $2700$,
$D^{+}_{sJ}(1D)$, with a mass around $2718$ MeV, and a width $73.0
\pm 10.4$ MeV.

{\Large \bf Acknowledgements}

This work was supported in part by the National Natural Science
Foundation of China (NSFC) under Grant No. 10875032, and in part by
SRF for ROCS, SEM.

\end{document}